\documentclass[aip,reprint,amsmath]{revtex4-1}
\usepackage{graphicx}% Include figure files
\usepackage{dcolumn}% Align table columns on decimal point
\usepackage{bm}% bold math
\usepackage[english]{babel}

\usepackage{color}% color text

\newif\ifgraph

\graphtrue  
\numberwithin{equation}{section}

\begin{document}

\title{Nonlocality of relaxation rates in disordered landscapes}
\author{Yunyun Li}
\email{yunyunli@tongji.edu.cn}
\affiliation{Center for Phononics and Thermal Energy Science, School of Physics Science and Engineering, Tongji University, Shanghai 200092, People's Republic of China}
\affiliation{ China-EU Joint Lab for Nanophononics, Tongji University, Shanghai 200092, China}
\affiliation{ Shanghai Key Laboratory of Special Artificial Microstructure Materials and Technology, School of Physics Science and Engineering, Tongji University, Shanghai 200092, China}

\author{Debajyoti Debnath}
\affiliation{Department of Chemistry, Presidency University, Kolkata 700073, India}

\author{Pulak K. Ghosh} 
\affiliation{Department of Chemistry, Presidency University, Kolkata 700073, India}

\author{Fabio Marchesoni}
\affiliation{Center for Phononics and Thermal Energy Science, School of Physics Science and Engineering, Tongji University, Shanghai 200092, People's Republic of China}
\affiliation{ China-EU Joint Lab for Nanophononics, Tongji University, Shanghai 200092, China}
\affiliation{Dipartimento di Fisica, Universit\`{a} di Camerino, I-62032 Camerino, Italy}

\date{\today}

\begin{abstract}
We investigate both analytically and by numerical simulation the relaxation of an overdamped Brownian particle in a 1D multiwell potential. We show that the mean relaxation time from an injection point inside the well down to its bottom is dominated by statistically rare trajectories that sample the potential profile outside the well. As a consequence, also the hopping time between two degenerate wells can depend on the detailed multiwell structure of the entire potential. The nonlocal nature of the transitions between two states of a disordered landscape is important for the correct interpretation of the relaxation rates in complex chemical-physical systems, measured either through numerical simulations or experimental techniques.
\end{abstract}
%\pacs{05.60.-k, 05.40.-a, 68.43.Mn} 
\maketitle

%%%%%%%%%%%%%%%%%%%%%%%%%%%%%%%%%%%%%%%%%%%%%%%%%%%%%%%%%%%%%%%%%%%%%%%%%%%%%%%%%%%%%%%%%%%%%%%%%
\section{Introduction}
\label{introduction}

The problem of time relaxation around a local minimum of a free-energy landscape is ubiquitous in chemical physics. In fact, the landscape  picture  \cite{goldstein} assumes a natural separation  of  low-temperature  molecular  motion sampling  distinct potential energy minima, and vibration within a minimum. The manner in which a disordered material samples its landscape as a function of temperature thus provides information  on  its  long-time relaxation properties. The energy landscape paradigm has been successfully applied to protein folding \cite{frauen}, the mechanical properties of glasses \cite{malandro}, and the dynamics of supercooled liquids \cite{sastry}.

In this context, Adam-Gibbs' formula \cite{adam} suggests a phenomenological connection between kinetics and thermodynamics in disordered systems, that is, $\tau= A\exp(B/Ts_c)$, where $\tau$ is a relaxation time, $A$ and $B$ are two phenomenological constants, and $s_c$ is a configurational entropy factor related to the number of minima of the system's multidimensional energy surface. For instance, at low enough temperatures the system becomes stuck in a single  minimum,  the  depth  of  which  increases  as  the  cooling  rate decreases: this describes a glass transition. In this context, of prominent interest is the case of relaxation between two degenerate free-energy minima separated by an (almost) symmetric activation barrier. In the current literature this is referred to as the Kramers' problem \cite{borkovec}. In calculating the average transition time between two such states, one typically ignores the presence of other possible less stable (more energetic) states in the free-energy landscape \cite{tang}. We show that statistically rare trajectories that connect two such degenerate states only after entering another neighboring state, are responsible for an increase of the relevant mean transition time, sometimes by orders of magnitude. The consequence is that in order to ignore the contribution of slowly meandering trajectories and keep using the results of standard Kramers' theory, one has to restrict the system's phase-space volume defining the free-energy stable states.

Our conclusion has an immediate counterpart and, hopefully, application in the strategies of path sampling for the numerical investigation of complex systems \cite{ferrario,bowman}. For instance, an unfolded protein can explore thousands of intermediate structures (conformations) before reaching a long-lived (stable) folded conformation. The most numerically efficient approach to investigate this process involves simulating protein folding with molecular dynamics for a relatively short time, and then analyzing the resulting trajectories to extract a coarse-grained Markov state model (MSM). An MSM consists of an appropriate choice of long-lived clustered conformational states and the transition rates between them. To create an MSM, one runs molecular dynamics simulations to determine how frequently a protein changes from one state to another, and clusters intermediate structures based on kinetic proximity (e.g., how energetically easy is switching from one structure to another). The transition rates are typically determined by averaging the time the protein takes to switch between any two states encoded in the MSM. Due to the coarse-grained nature of the MSM, a continuous trajectory connecting a pair of sampled states might well enter first the phase-space basin belonging to another state without being trapped there. This occurrence, though unlikely, may dramatically affect the corresponding transition time. How to correctly generate the reactive trajectories representing a specific transition of interest for the MSM is an issue of ongoing research.

The contents of this paper is organized as follows. In Sec. \ref{bistable} we first simulate the relaxation of an overdamped Brownian particle in a 1D potential well. We determine both numerically and analytically the mean first-passage time (MFPT) for the particle to reach the bottom of the well from an injection point inside it. We show that when the injection point rests inside the well, but higher than the bottom of another adjacent well, then the rare trajectories crossing the barrier separating the two wells become dominant and, on lowering the noise level, the MFPT increases exponentially. Some of the results presented here have been independently derived in Ref. \cite{AA} for discrete stochastic models of biological interest. In Sec. \ref{dominant} we interpret this effect by distinguishing between two types of trajectories, the most probable trajectories pointing from the injection point straight down to the well bottom, and the rare trajectories overcoming the barrier into the side well. The distribution density of the relaxation times allows a clear-cut distinction between these two types of trajectories. In Sec. \ref{multiwell} we extend our analysis to the case of multiwell potentials and conclude that the MFPT inside a well is dominated by barrier-crossing anytime the particle's injection point rests above the level of the lowest lying among all adjacent wells (Sec. \ref{threewellNondeg}). Finally, we consider the case of the hopping process between two degenerate minima of the potential and discuss how the MFPT over the barrier separating them can depend on the level of the injection point and, therefore, on the multistable structure of the entire potential (Sec. \ref{threewellDeg}).  In Sec. \ref{conclusions} we draw some concluding remarks regarding the impact of this effect on the interpretation of actual relaxation measurements.

%%%%%%%%%%%%%%%%%%%%%%%%%%%%%%%%%%%%%%%%%%%%%%%%%%%%%%%%%%%%%%%%%%%%%%%%%%%%%%%%%%%%%%%%%%%%%%%%%
\section{Relaxation times in a bistable potential}
\label{bistable}

We start introducing two categories of trajectories a 1D system may take while relaxing toward a stable state. Broadly speaking, we distinguish between \textit{regular} trajectories, the most probable and typically the shortest ones, given certain initial conditions, and a subset of \textit{dominant} trajectories, which one determines with
reference to the observable being measured. The most probable transition trajectories in a 1D system has been classified by analyzing the (local) minima of the relevant action integrals \cite{path1,path3}. Here, we are rather concerned with identifying the systems' trajectories that most contribute to the mean value of a specific observable of interest.

\begin{figure}
\begin{center}
\includegraphics[width=8.5truecm]{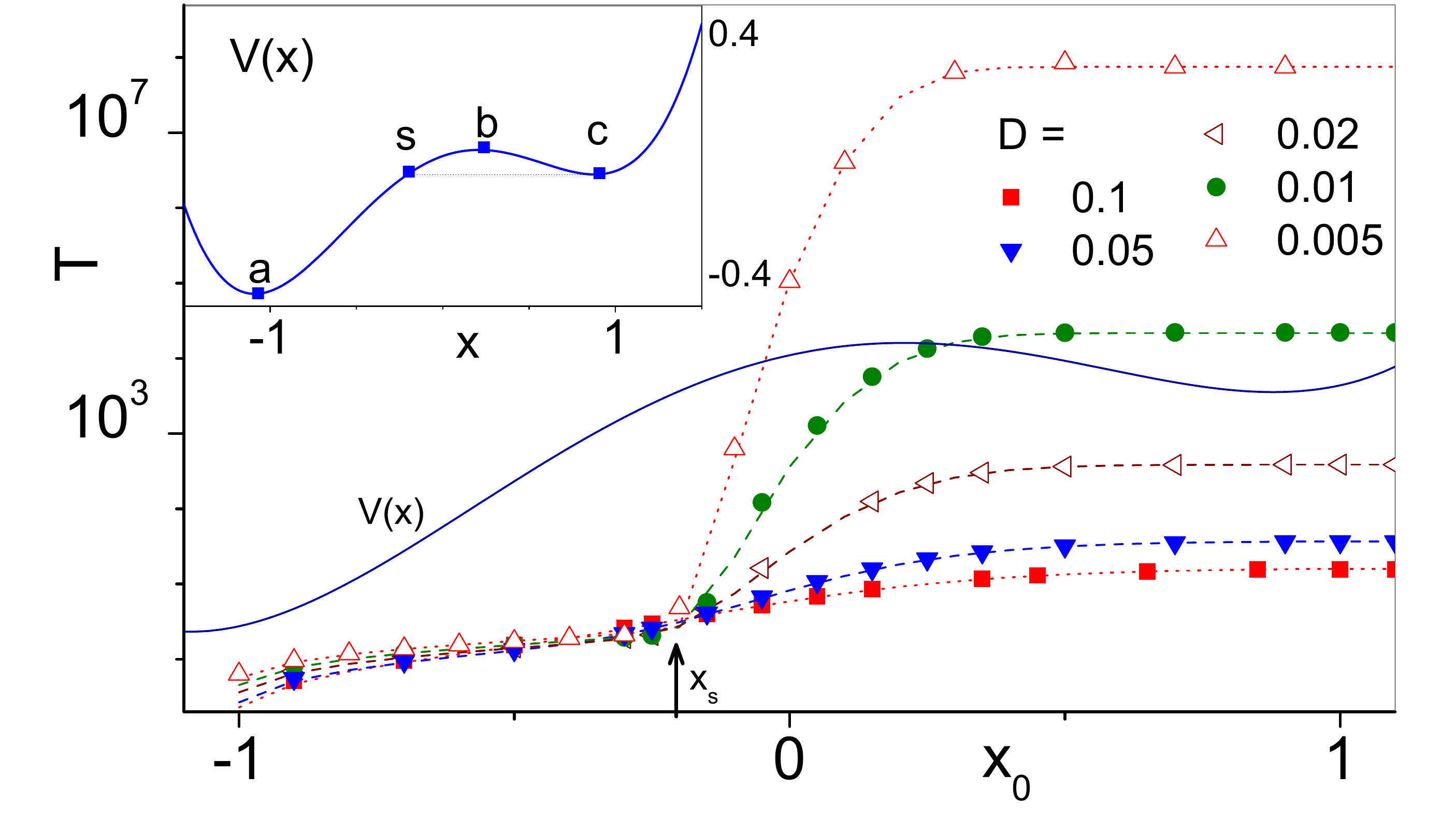}
\caption{(Color online) Mean first-passage time $T=T(a,x_0)$ vs. $x_0$ from numerical integration of the LE (\ref{2.1}) with the asymmetric bistable potential $V(x)=x^4/4-x^2+x/5$ (inset) and different noise strength, $D$. The potential minima are located at $x_a\approx-1.088$ and $x_c\approx0.879$, the barrier at $x_b\approx0.209$, and the crossover threshold, defined by $V(x_s)=V_c$, at $x_s\approx-0.204$. The dashed curves $T(a,x_0)$ were obtained by performing the double integral in Eq. (\ref{2.4}) for the appropriate $D$.\label{F2.1}}
\end{center}
\end{figure}

A study-case is represented by the transition times $t(a,x_0)$ of an overdamped Brownian particle obeying the Langevin equation (LE),
\begin{equation} \label{2.1}
\dot{x} =  - V'(x) + \xi (t),
\end{equation}
where $x(t)$ denotes the particle coordinate, $V(x)$ is a confining
multistable potential, and $\xi(t)$ models a stationary, zero-mean,
Gaussian noise source with autocorrelation function
\begin{equation} \label{2.2}
\langle \xi(t) \xi(0)\rangle = 2 D \delta(t).
\end{equation}
The particle will be injected at a given point $x_0$ and taken out
upon reaching the exit point $x_a$. To keep our notation as simple
as possible, we place the exit point at the bottom of a potential
well, termed well $a$, located on the left of the injection
point, i.e., $x_a<x_0$, see inset of Fig. \ref{F2.1}.  The time length of each trajectory is the observable of interest, $t(a,x_0)$.

The average transition time $T(a,x_0)\equiv \langle t(a,x_0)
\rangle$ for the particle to diffuse from $x_0$ to $a$, is given by
the well-known MFPT formula
\cite{stratonovich,gardiner,goel},
\begin{equation} \label{2.3}
T(a,x_0)=
\frac{1}{D}\int_{x_a}^{x_0}\frac{dy}{p(y)}\int_y^{\infty}p(z)dz,
\end{equation}
where $p(x)={\cal N}\exp[-V(x)/D]$ is the stationary probability
density of the process (\ref{2.1}). Note that for a confining
potential, $\lim_{x\to\pm\infty}p(x)=0$, i.e., $x\to \infty$ can be
treated as a reflecting boundary \cite{gardiner}

We specialize now Eq. (\ref{2.3}) to the case of an asymmetric bistable
potential. As illustrated in the inset of Fig. \ref{F2.1}, $x_b$ locates the top of the barrier, $b$, and $x_a$ and $x_c$ denote the bottom of the left, $a$, and right well, $c$, respectively, with $V_a < V_c$. Here and in the
following, we adopted the short-hand notation $V(x_a)=V_a$,
$V(x_b)=V_b$, $V(x_c)=V_c$, $V(x_0)=V_0$, and prime for an $x$ derivative, $(\dots)'=d(\dots)/dx$. The threshold $x_s$ is the point on the r.h.s. of $a$ that has the same potential energy as the bottom of well $c$; for the asymmetric double-well potential of Fig. \ref{F2.1}, $V(x_s)=V_c$ with $x_s>x_a$.

We then estimate  the MFPT (\ref{2.3}) in the weak noise limit,
$D<V_b-V_c$, for three different ranges of the injection point,
$x_0$:

\textit{(i) out-of-well, $x_0>x_b$.} The functions $p(x)$ and
$p^{-1}(x)$ are sharply peaked, respectively, around points $x_a$
and $x_c$ and around point $x_b$. As a
consequence, for $x_0>x_b$ the nested integrals (\ref{2.3})
factorize, that is,
\begin{equation} \label{2.4}
T(a,x_0)=
\frac{1}{D}\int_{x_a}^{x_0}\frac{dy}{p(y)}\int_{x_b}^{\infty}p(z)dz.
\end{equation}
In the limit of weak noise \cite{gardiner} $p(z)/p(y)\simeq
\exp[(V_b-V_c)/D -|V_b''|(x-x_b)^2/2D-V_c''(x-x_c)^2/2D]$, so that the
integrals (\ref{2.4}) can be approximated to
\begin{equation} \label{2.5}
T(a,x_0)=
\frac{2\pi}{\sqrt{|V_b''|V_c''}}\exp\left(\frac{V_b-V_c}{D}\right).
\end{equation}
This is the well-known Kramers' formula, $T_K(a,c)$, for the escape
time out of well $c$. Here, according to our notation, all escape trajectories are regular and the ensuing (almost $x_0$ independent) relaxation time is characterized by the
slow relaxation process $x_c \to x_a$.

\textit{(ii) barrier well region, $x_s<x_0<x_b$.} For this choice
of the injection point, the first integrand (\ref{2.4}) can be
approximated to $p^{-1}(y)\simeq \exp[V_0/D + V_0'(y-x_0)/D]$;
hence
\begin{equation} \label{2.6}
T(a,x_0)= \frac{1}{|V_0'|}\sqrt{\frac{2\pi
D}{V_c''}}\exp\left(\frac{V_0-V_c}{D}\right).
\end{equation}
Here we took the absolute value of $V_0'$ only for the sake of
generality. This result is suggestive: Although the particle was
injected directly in well $a$, still it takes an exponentially long
average time to reach its bottom, $x_a$.
Moreover, in contrast with Kramers' time of Eq. (\ref{2.5}), $T(a,x_0)$
appears to depend on how high the injection point lies with respect
to the minimum, $V_c$, of the side-well $c$. As discussed in Sec.
\ref{dominant}, the MFPT (\ref{2.6}) is indeed dominated by the rare
trajectories that cross over into well $c$ before being absorbed at
$x_a$.

\textit{(iii) bottom well region, $x_0<x_s$.} As $x_0$ approaches
the exit point, one can easily take the $x_0 \to x_a$ limit of the
double integral (\ref{2.3}), thus obtaining the logarithmic law,
\begin{equation} \label{2.7}
T(a,x_0)=\frac{1}{2V''_a}\left [2 \ln 2 +\gamma+\ln \left(\frac{V_0-V_a}{D} \right )\right ],
\end{equation}
where $\gamma\simeq 0.577$ is the Mascheroni's constant. This is the short MFPT one would expect on account of the sole
regular trajectories of the relaxation process. Indeed, such
trajectories run straight downhill from $x_0$ subject to weak noise
fluctuations, whose effect grows appreciable only close to the exit
point, $x=x_a$.

Our analytical estimates (\ref{2.5})-(\ref{2.7}) reproduce well the
three different regimes of the $T(a,x_0)$ curves of Fig.
\ref{F2.1}, obtained by numerically computing the double integral
(\ref{2.3}) for very small $D$ values. The crossover between the logarithmic
(\ref{2.7}) and the exponential branch (\ref{2.6}) of $T(a,x_0)$ is
fairly sharp, because the exponential in Eq. (\ref{2.6})
abruptly vanishes for $x_0<x_s$ and $|V_0-V_c|\ll D$.

In passing we notice that our approximations (\ref{2.5}) and
(\ref{2.6}) coincide (apart from minor typographical errors) with
the first two MFPT's reported in Eq. (33) of Ref.\cite{AA} for
Schl\"ogl's model in the large size system limit. Our derivation is
much simpler, indeed, but restricted to the case of continuous
stochastic transition processes.

Finally, the results of this section can be readily extended to the
case when the side-well $c$ is deeper than the exit well, $V_c<V_a$.
Only approximation (\ref{2.7}) needs to be modified as the
probability density, $p(x)$, in the exit well gets exponentially suppressed.
As a consequence, the right hand side of Eq. (\ref{2.7}) must be multiplied by the additional factor $\exp[(V_a-V_c)/D]$. This means that, since no
threshold $x_s$ could be defined, the average transition time is
exponentially long for any in-well injection point, namely,
$T(a,x_0)\propto \exp[(V_0-V_c)/D]$ for $x_a < x_0 < x_b$.

%%%%%%%%%%%%%%%%%%%%%%%%%%%%%%%%%%%%%%%%%%%%%%%%%%%%%%%%%%%%%%%%%%%%%%%%%%%%%%%%%%%%%%%%%%%%%%%%%
\section{The role of the dominant trajectories}
\label{dominant}

As anticipated in the foregoing section, the results of Eqs. (\ref{2.5}) and (\ref{2.7}) lend themselves to a simple interpretation in terms of regular trajectories. For $x_0 > x_b$ the particle is initially placed in
the side-well $c$, so that it, first, relaxes around the local
$p(x)$ maximum at $x_c$ and, then, escapes into well $a$ by
overcoming the barrier $b$; as a consequence $T(a,x_0)$ is quite
insensitive to the injection point $x_0$. For $x_a <x_0<x_s$ the
particle tends to roll downhill toward the exit point $x_a$,
corresponding to the absolute maximum of $p(x)$, with a short
average transition time proportional to the logarithm of the initial displacement, $x_0-x_a$.

\begin{figure}
\begin{center}
\includegraphics[width=8.0truecm]{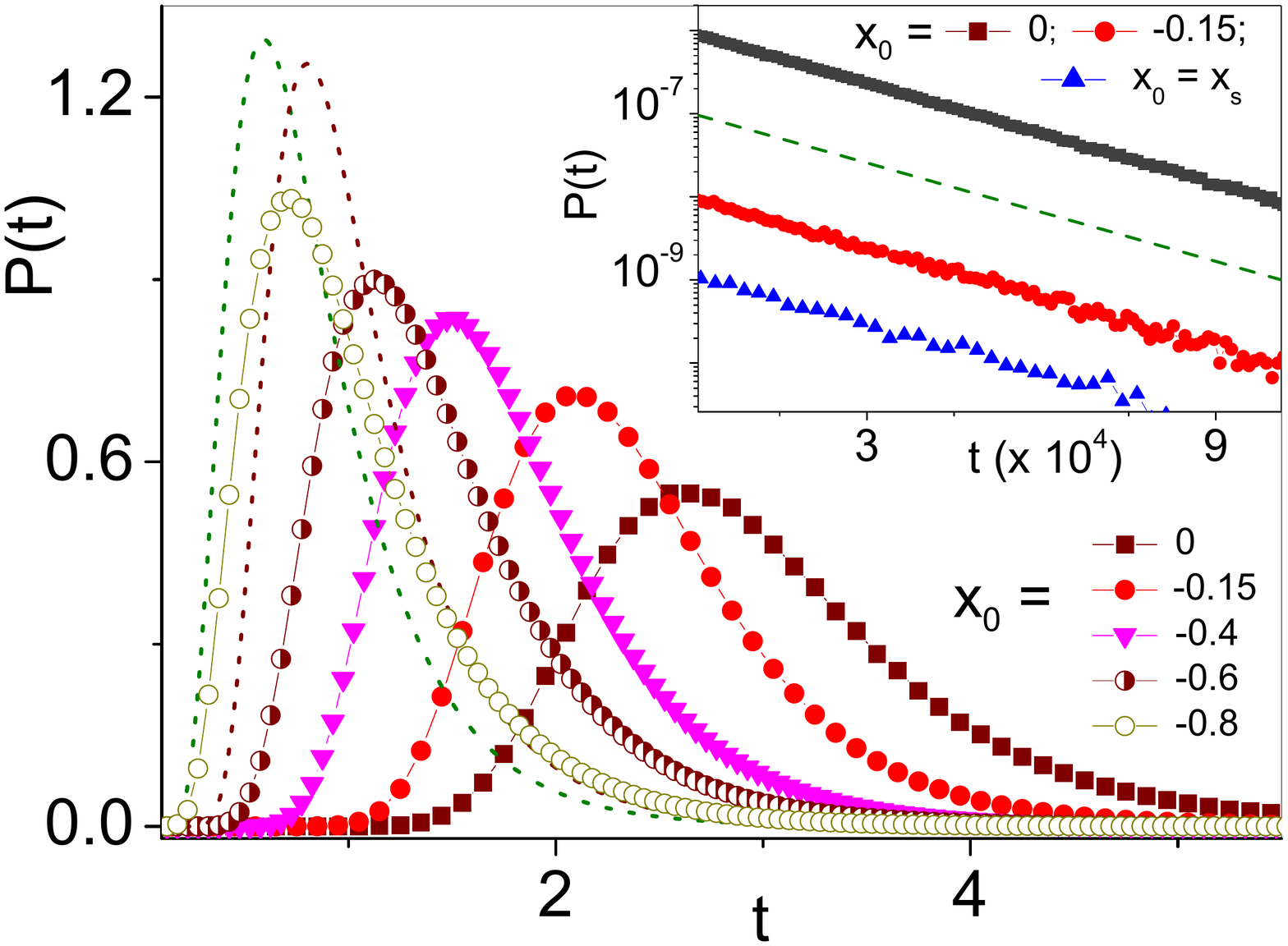}
\caption{(Color online) Distribution densities, $P(t)$, of the transient
times $t=t(a,x_0)$ obtained by numerically integrating the LE
(\ref{2.1}) for the asymmetric bistable potential of Fig. \ref{F2.1}
with $D=0.01$ and different $x_0$. The dashed curves represent the harmonic approximation $P_s(t)$ of Eq. (\ref{PTharm}) for the two $x_0$ closest to $x_a$, see text. Inset:  semi-logarithmic plot of $P(t)$ vs $t$ for three values of $x_0$ \ and $D=0.01$. The three data sets are closely fitted by the function $P_l(t)$ in Eq. (\ref{3.4}). The dashed line with $T_K(x_a,x_0)$ has been drawn to guide the eye. \label{F3.1}}
\end{center}
\end{figure}

The transitions that start out in the barrier region $x_s < x_0 < x_b$
are qualitatively different. As the injection point lies inside well
$a$, the trajectories oriented toward the exit point are still the
most probable, or, stated otherwise, they represent the process' \textit{regular} trajectories, as expected.
Nevertheless, the particle can diffuse from $x_0$ over
the barrier into well $c$ with small but finite probability. Following Refs.
\cite{gardiner,goel}, we can estimate the splitting probability
$\pi(a,x_0)$ for the particle to exit at $a$ without first
reaching $c$, and $\pi(c,x_0)$ for the particle to fall into well $c$
before being absorbed at $a$,
\begin{equation} \label{3.1}
\pi(c,x_0)= 1-\pi(a,x_0),
\end{equation}
\begin{equation} \label{3.2}
\pi(a,x_0)= \int_{x_0}^{x_c} \frac{dy}{p(y)} {\large /}\int_{x_a}^{x_c} \frac{dy}{p(y)}.
\end{equation}
For weak noises and $x_0$ not too close to the extrema $x_a$ and $x_b$, the integral (\ref{3.2}) can be approximated to \cite{gardiner}
\begin{equation} \label{3.3}
\pi(c,x_0)\simeq \frac{1}{|V_0'|}\sqrt{\frac{D|V_b''|}{2
\pi}}\exp\left(\frac{V_0-V_b}{D}\right).
\end{equation}
\begin{figure}
\begin{center}
\includegraphics[width=3.5in]{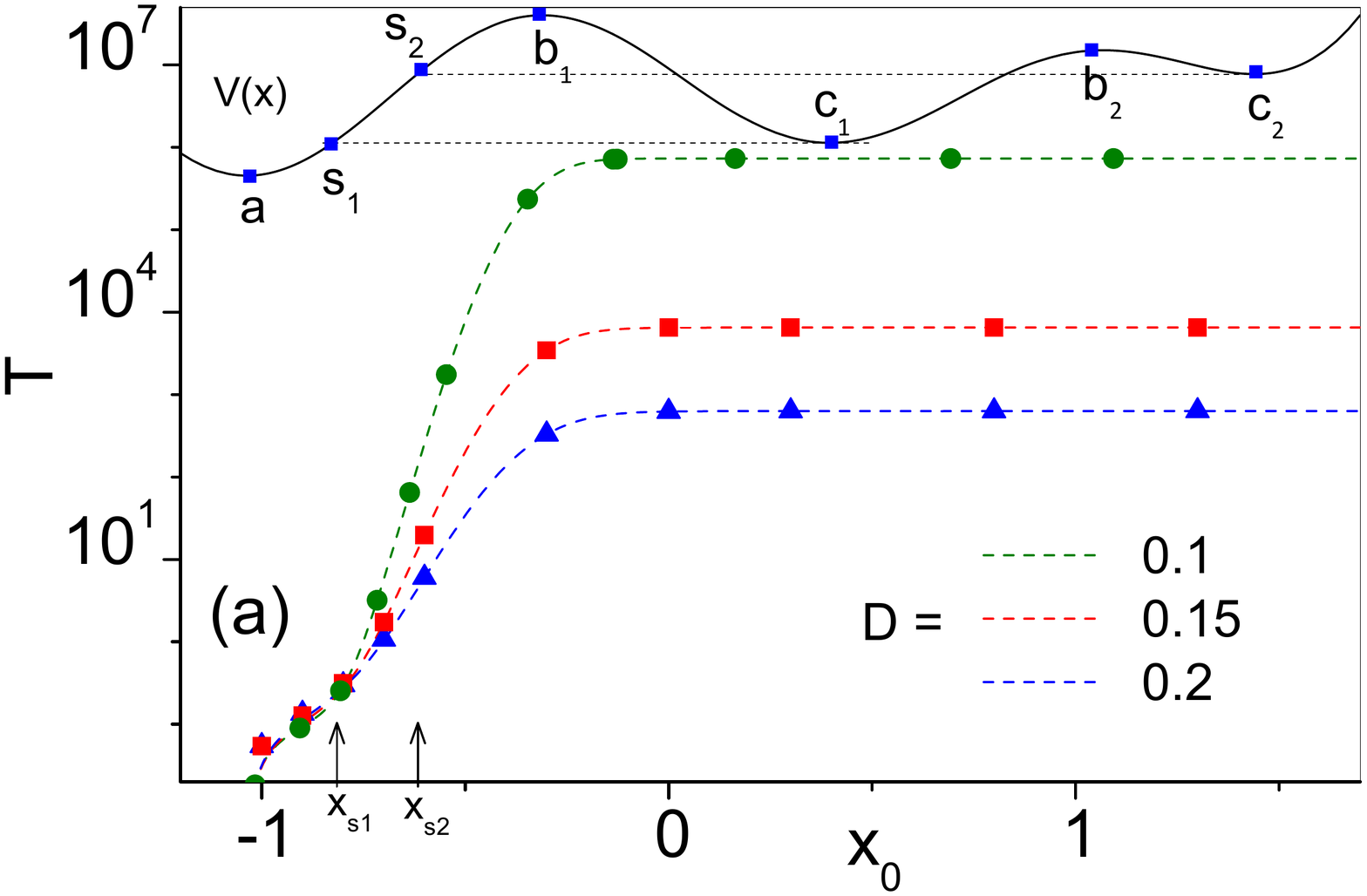}
\includegraphics[width=3.5in]{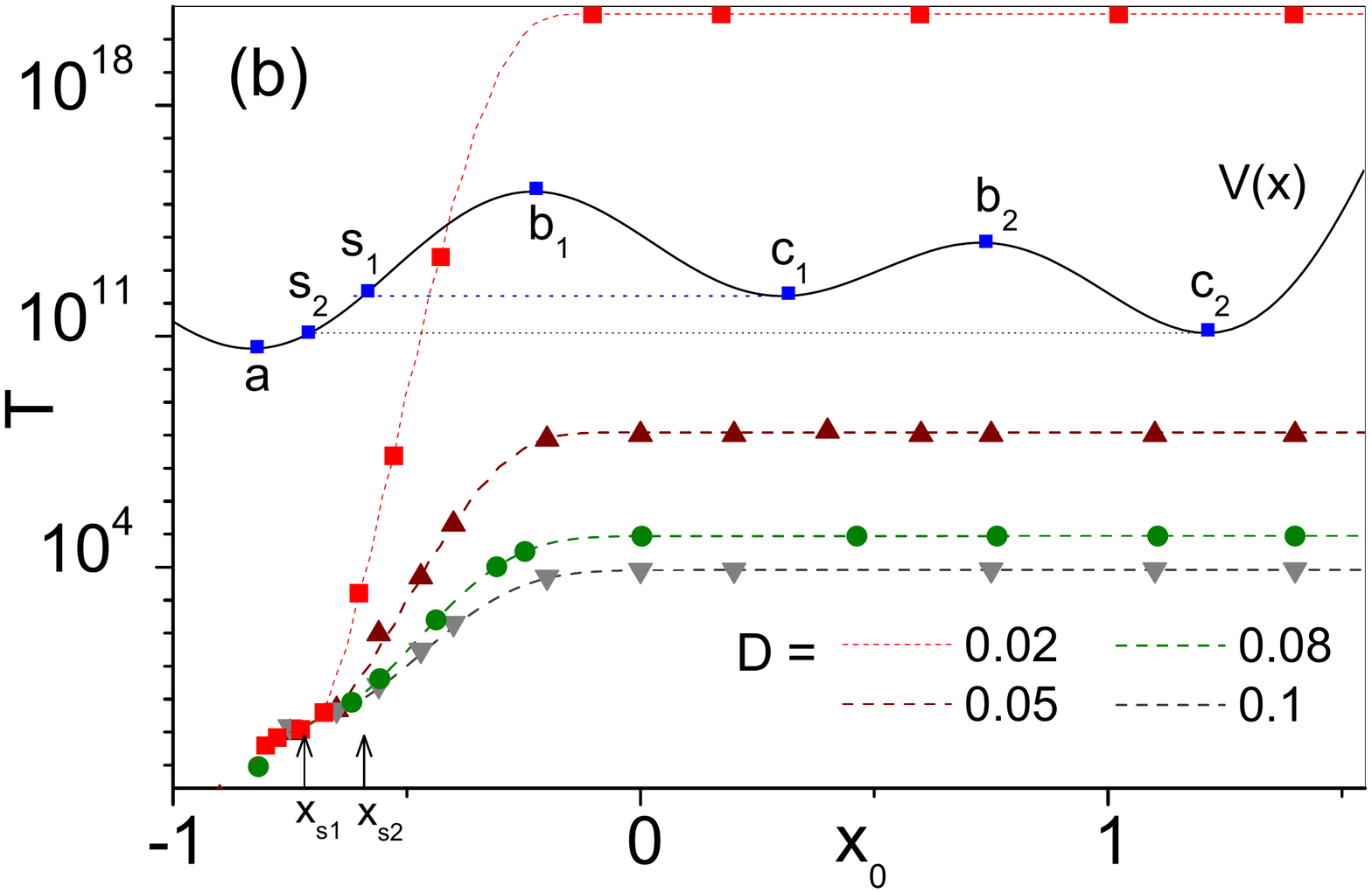}
\caption{(Color online) Mean first-passage time $T=T(a,x_0)$ vs. $x_0$ in the asymmetric three-well potentials (a)
$V(x)=x^4/2-x^2-0.7\sin(4.5x)$ and (b) $V(x)=x^4/2-x^2-0.35\sin(6x)$ for different $D$. Note that in (a) $V_{c_1}<V_{c_2}$ and in (b) $V_{c_1}>V_{c_2}$; in both cases $V_a<V_{c_i}$, $i=1,2$. The data points are the result of the numerical integration of LE (\ref{2.1}) for the relevant choices of $V(x)$ and $D$; the dashed curves are the corresponding analytical expressions of Eq. (\ref{2.4}).  \label{F4.1}}
\end{center}
\end{figure}
Although the typical trajectories are by far the most probable --
being $\pi(a,x_0)\simeq 1$, -- still their contribution to the
average transition time $T(a,x_0)$ is negligible, as they reach the
exit point in a quite short time, see Eq. (\ref{2.7}). By contrast, the
barrier crossings may well be very unlikely -- being $\pi(c,x_0)$
exponentially small, -- but the particle, after falling into well
$c$, takes an exponentially long time of the order of $T_K(a,c)$ [see Eq. (\ref{2.3}) for $x_0=c$], to
recross into well $a$. The contribution to $T(a,x_0)$ from such rare
trajectories amounts to $\pi(c,x_0)T_K(a,c)$, that is, to our
estimate in Eq. (\ref{2.6}). In conclusion, as long as we
characterize the relaxation in the overdamped potential $V(x)$ by measuring the exit times, $t(a,x_0)$, the otherwise sporadic trajectories crossing the barrier may become \textit{dominant}, depending on the injection point. Of course, this argument only applies for small, but finite noise strengths, i.e., $D \to 0+$, whereas in the noiseless regime, $D=0$, there exists only one allowed deterministic trajectory running downhill from $x_0$ to $x_a$ for any $a<x_0<x_b$.

In real or numerical experiments one can easily sample relaxation
trajectories from $x_0$ to $x_a$ and distribute them according to
their temporal length, $t(a,x_0)$. Based on the argument above,
where the regular trajectories are regarded as much faster than the
dominant ones, the $t$-distribution density, $P[t(a,x_0)]$, can be
separated into two distinct terms, i.e.,
$\pi(a,x_0)P_s(t)+\pi(c,x_0)P_l(t)$. For the sake of a
comparison with actual data, in the regime of weak noise one
can introduce the approximations,
\begin{equation} \label{3.4}
P_l(t)\simeq \frac{T(a,x_0)}{T_K^2(a,c)}\,\exp[-{t}/{T_K(a,c)}],
\end{equation}
for the long exit times of the statistically rare trajectories crossing the barrier, and
\begin{equation} \label{PTharm}
P_s(t)\simeq -\frac{2}{\sqrt{\pi}} \frac{d}{dt} \left[\frac{V_0-V_a}{D(t)} \right]^{\frac{1}{2}} \cdot \exp[-({V_0-V_a})/{D(t)}],
\end{equation}
with $D(t)=e^{2V_a''t}-1$, for the intrawell relaxation trajectories. Our expression for $P_s(t)$ holds good for the harmonic approximation of the potential well $a$, that is, by setting $V(x) = V_a+(1/2)V_a''(x-x_a)^2$ and ignoring all anharmonic terms of the third order and higher. It was derived by standard MFPT methods \cite{gardiner} and can be reformulated to match earlier solutions for $t$-distribution in a harmonic well \cite{szabo,berne}. In Eq. (\ref{3.4}) for $P_l(t)$, we approximated the probability of barrier crossing as $\pi(c,x_0)\simeq T(a,x_0)/T_K(a,c)$, and made use of the well-established exponential distribution for Kramers' escape times from $c$ back to $a$ \cite{borkovec,gardiner,goel}.

In Fig. \ref{F3.1} we display the outcome of an extensive numerical
simulation of the exit process, Eq. (\ref{2.1}), for the potential of Fig.
\ref{F2.1} and different values of $D$. As the injection point is shifted past the threshold $x_s$, also the relaxation time distributions change
abruptly. An exponential tail associated with the dominant trajectories becomes visible for $x_0\geq x_s$ (inset); as predicted in Eq. (\ref{3.4}), such
a tail has a small amplitude of the order of $T(a,x_0)/T_K^2(a,c)$ and decays slowly with time constant $T_K(a,c)$. The distributions of the short relaxation
times due to the regular trajectories,  main panel, are reminiscent of the $t$-distributions in a harmonic well, $P_s(t)$ of Eq. (\ref{PTharm}). However, the agreement gets quantitatively close only when $x_0$ approaches $x_a$, the convergence being rather slow. We attributed this inconvenience to the spatial asymmetry of well $a$. Moreover, we remark that the average $\langle t(a,x_0)\rangle$ taken over the regular trajectories only, namely by using the approximate distribution density Eq. (\ref{PTharm}), is a monotonic decreasing function of $D$; for vanishingly small $D$ values it comes close to the predicted estimate in Eq. (\ref{2.7}).

%%%%%%%%%%%%%%%%%%%%%%%%%%%%%%%%%%%%%%%%%%%%%%%%%%%%%%%%%%%%%%%%%%%%%%%%%%%%%%%%%%%%%%%%%%%%%%%%%
\section{Generalization to multiwell potentials}
\label{multiwell}

The results of Sec. \ref{bistable} can be extended to
study transitions in multiwell potentials, as well. However, the
algebraic manipulations on the MFPT (\ref{2.3}) can become more
complicated due to the multi-peaked structure of the functions $p(x)$
and $p^{-1}(x)$. Luckily, to gain a better understanding of the role of the dominant trajectories in the most general case of a disordered potential, it suffices to analyze in some detail the three-well potentials, only. While any disordered potential can be regarded as an appropriate sequence of three-well potentials, it is clear that the relaxation properties discussed below only apply in the limit of infinite observation times, where the diffusing particle is allowed to explore the entire potential profile. Shorter observation times would necessarily restrict our analysis to the portion of the potential profile actually accessed by the particle.

\subsection{Nondegenerate three-well potentials}
\label{threewellNondeg}

Let us imagine to add a third well to the potential plotted in Fig.
\ref{F2.1}. If we agree on that the exit well must be at the bottom of the lowest one, then two geometries are possible, as illustrated in Fig. \ref{F4.1}.
Let $c_1$ and $c_2$ denote, respectively, the first and the second
well to the right of well $a$, with barriers $b_1$ and $b_2$
separating the three wells. As for both wells $V_{c_i}>V_a$, with
$i=1,2$, the equations $V(x_{s_i})>V_{c_i}$ may define two thresholds,
$x_{s_i}$, with $a<x_{s_i}<x_{b_1}$. As a consequence, the barrier region
of well $a$ is delimited from below by the threshold
$\overline{x}_s\equiv \min\{x_{s_i}\}$, that is, it starts at the
level of the lower side-well $c_{i}$ -- see the geometric constructions
in panels (a) and (b).

Now, the question rises naturally whether, in the presence of two (or
more) thresholds, the slope of $T(a,x_0)$ changes at each of them, and where such changes are possibly the most pronounced. The
answer is illustrated in the two panels of Fig. \ref{F4.1}, where the MFPT
(\ref{2.3}) has been plotted over an $x_0$ range comprising both
$x_{s_i}$:  On reducing the noise intensity, a sharp crossover
between a logarithmic and an exponential $x_0$ dependence emerges in the
neighborhood of $\overline{x}_s$, whereas no substantial MFPT change  can be associated with the other threshold. This conclusion can be
confirmed qualitatively by extending the semi-quantitative approach of Sec.
\ref{bistable} to both potentials of Fig. \ref{F4.1}. In the
barrier region, the average transient time $T(a,x_0)$ is dominated
by the lower side-well $c_{i}$; the dominant trajectories cross one
or two barriers, depending on which side-well is deeper.
Accordingly, in the barrier region $\overline{x}_s<x_0<x_{b_1}$, the
curve $T(a,x_0)$ grows proportional to
$\exp([V_0-V(\overline{x}_s)]/D)$. Note that in view of the
remark at the bottom of Sec. \ref{bistable}, should one side-well $c_i$ sit
lower than well $a$, then such an exponential dependence would apply throughout the entire range $x_a<x_0<x_{b_1}$ and no logarithmic-to-exponential crossover would occur.

\subsection{Degenerate three-well potentials}
\label{threewellDeg}

\begin{figure}
\begin{center}
\includegraphics[width=7.5truecm]{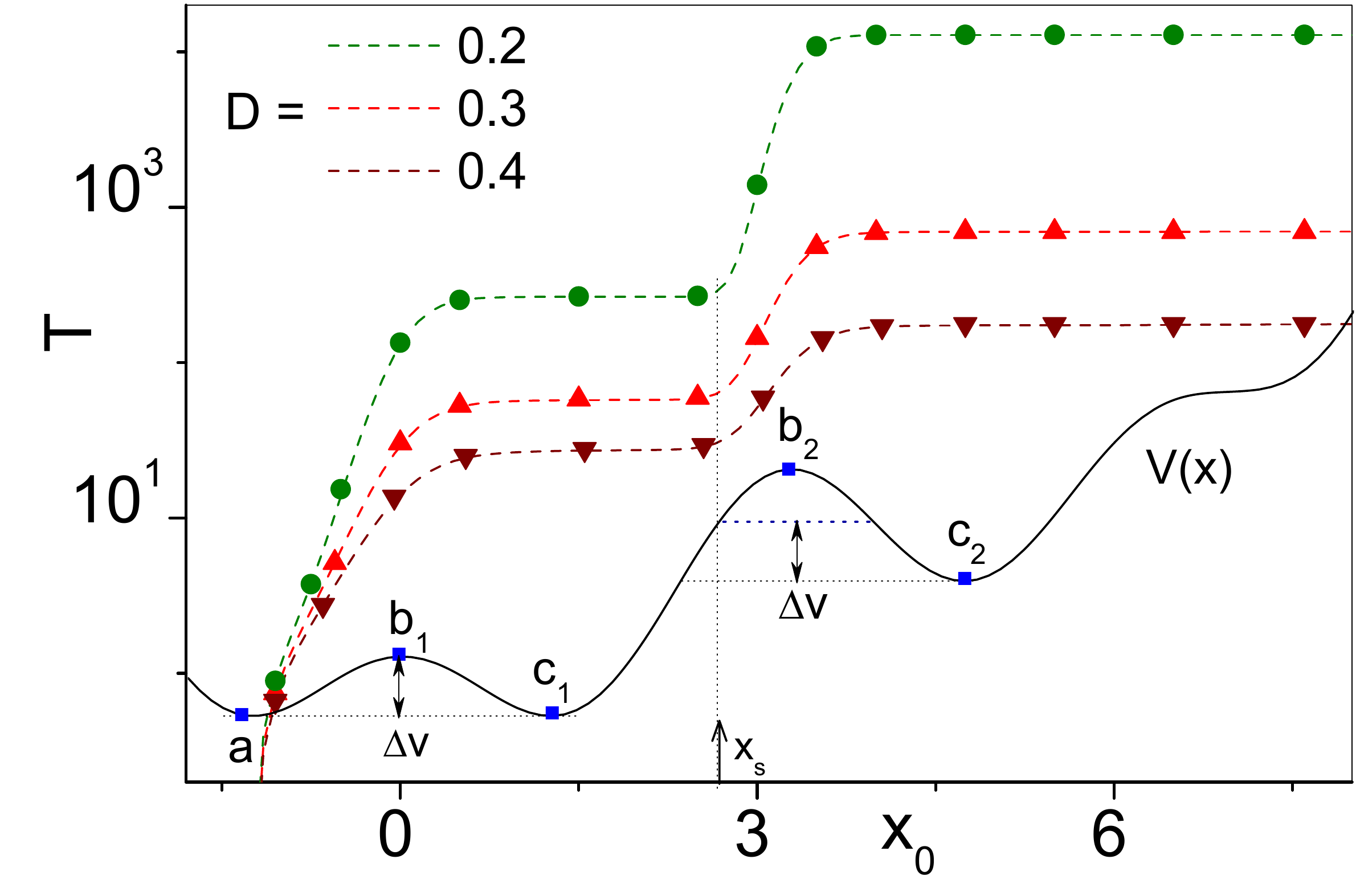}
\caption{(Color online)  Mean first-passage time $T=T(a,x_0)$ vs. $x_0$ in the degenerate three-well potential $V(x)=\cos 2x-\cos x++0.1x$ for different $D$; simulation (symbol) versus analytical results (dashed curves). Note that  $V_{a}=V_{c_1}$ and the threshold $x_0=x_s$ is defined by the condition $V(x_s)=V_{c_2}+\Delta V$, where $\Delta V=V_{b_1}-V_{c_1}$  (see text). The dashed curves were obtained by numerically integrating Eq. (\ref{2.4}) for the three-well potential. \label{F4.2}}
\end{center}
\end{figure}

We consider now the special case of a three-well potential with two degenerate lower minima, say, in $x_{a}$ and $x_{c_1}$, see Fig. \ref{F4.2}. This means that wells $a$ and $c_1$ are equally deep, while the third well sits higher up, that is, $V_a=V_{c_1}<V_{c_2}$. Then, the process (\ref{2.1}) models the relaxation occurring between two degenerate states, a mechanism often invoked in the chemical physical literature. As discussed in Sec. \ref{introduction}, for low noise levels this problem is commonly addressed by ignoring the presence of more energetic states in the neighborhood. However, the remarkable dependence of $T(x_0,a)$ on the injection point, $x_0$, shown in figure, suggests a different picture. As long as $x_0$ is confined around the bottom of well $c_1$, the MFPT from $x_0$ to $x_a$ is almost independent of $x_0$ and  well reproduced by the Kramers' rate of Eq. (\ref{2.5}) upon replacing $x_0$ with $c_1$, and $b$ with $b_1$. In this case the role of well $c_2$ is irrelevant. However, on moving $x_0$ to the right of a certain threshold $x_s$, $T(x_0,a)$ suddenly jumps up to a much higher value, insensitive to any further increase of $x_0$.

The location of the threshold point $s$ and the magnitude of the MFPT jump can be explained as follows. We assume that the lower $T(x_0,a)$ plateau for $x_{b_1}<x_0<x_{s}$ is due to the regular trajectories crossing from $c_1$ to $a$ directly over barrier $b_1$ and, therefore, proportional to $\exp(V_{b_1}-V_{c_1})$, whereas the higher plateau must come from those rare trajectories that cross first barrier $b_2$ to the right, with probability proportional to $\exp[-(V_{b_2}-V_{0})]$. The time they take to cross back from well $c_2$ to well $c_1$ (and then to well $a$) is a Kramers's time proportional to $\exp(V_{b_2}-V_{c_2})$. Therefore, their weighted contribution to the MFPT is proportional to $\exp(V_{0}-V_{c_2})$ and, most remarkably, supersedes the contribution from the regular trajectories for $V_0-V_{c_2}> V_{b_1}-V_{c_1}$. Accordingly, $x_s$ is determined by choosing $V(x_s)=V_{c_2}+\Delta V$, where $\Delta V=V_{b_1}-V_{c_1}$ is the barrier height separating wells $a$ and $c_1$ -- see the geometric construction in Fig. \ref{F4.2}.

As long as $V(x_s) < V_{b_2}$, the threshold $x_s$ is well defined. Therefore, there can exist a barrier region inside well $c_1$, $x_s < x_0 < x_{b_2}$, such that the relaxation trajectories creeping  into well $c_2$ are indeed {\it dominant}. The corresponding $t$-distributions are well fitted by double exponential functions (not shown) with decay constants equal to the two plateau values of the curves $T(x_0,a)$ versus $x_0$.

\section{Conclusions}
\label{conclusions}

Many systems in condensed matter are described by an overdamped particle that diffuses on a disordered energy landscape of appropriate dimensionality, without ever reaching a proper equilibrium state (glassy materials are a good example). The physical chemical properties of these systems are often interpreted in terms of the relaxation rates inside single locally stable states or between pairs of locally stable states. However, determining such rates experimentally, through microscopic techniques, or even numerically, may prove a moot problem. As discussed in Secs. \ref{bistable} and \ref{multiwell}, the investigator who intends to proceed by weakly exciting the system out of its locally stable state and then letting it relax back to it, may encounter the difficulty of establishing whether the measured relaxation time depends on the presence of other metastable states. This difficulty can be circumvented by a more restrictive definition of locally stable state.

Our analysis clearly shows that in 1D the relaxation times within a single potential well or between degenerate wells can be determined by ignoring additional potential wells only under the condition that the energy of what we call the injection point is sufficiently close to the energy of the well bottom. How close, it depends on the actual distribution of the wells along the potential landscape. Indeed, the critical threshold is determined by the lowest lying well, an information usually unavailable to the investigator. Therefore, above a certain (but unknown) threshold of the injection energy, the measured relaxation times exhibit a marked {\it nonlocal} dependence on the global potential profile. Such a nonlocal effect is due to the contribution from  slower, though rare, relaxation trajectories, which explore the potential landscape surrounding the well(s) of interest. Their presence can be appreciated, for instance, by looking at the distribution of the relevant relaxation times, though at the expense of much longer observation times.

The present analysis was restricted to 1D potentials for the sake of clarity, thus making our presentation hopefully easier to follow and affording higher numerical statistics. Its extension to potentials in two and even higher dimensions confirms the overall picture summarized here and is presently matter of further investigation.

%%%%%%%%%%%%%%%%%%%%%%%%%%%%%%%%%%%%%%%%%%%%%%%%%%%%%%%%%%%%%%%%%%%%%%%%%%%%%%%%%%%%%%%%%%%%%%%%%
\

\section*{Acknowledgements}
We thank RIKEN's RICC for computational resources.
Y. Li is supported by the NSF China under grant No. 11505128.
P.K.G. is supported  by SERB Start-up Research Grant (Young Scientist) No. YSS/2014/000853 and the UGC-BSR Start-Up Grant No. F.30-92/2015

%%%%%%%%%%%%%%%%%%%%%%%%%%%%%%%%%%%%%%%%%%%%%%%%%%%%%%%%%%%%%%%%%%%%%%%%%%%%%%%%%%%%%%%%%%%%%%%%%

\end{document}